# Extended Kohler's Rule of Magnetoresistance in TaCo$_2$Te$_2$

Samuel Pate[1,2], Bowen Chen[3], Bing Shen[3], Kezhen Li[1,2], Xiuquan Zhou[1], Duck Young Chung[1], Ralu Divan[4], Mercouri G. Kanatzidis[1,5], Ulrich Welp[1], Wai-Kwong Kwok[1], and Zhi-Li Xiao[1,2,*]

[1]*Materials Science Division, Argonne National Laboratory, Argonne, IL 60439, USA*

[2]*Department of Physics, Northern Illinois University, DeKalb, IL 60115, USA*

[3]*School of Physics, Sun Yat-sen University, Guangzhou 510275, China*

[4]*Center for Nanoscale Materials, Argonne National Laboratory, Argonne, IL 60439, USA*

[5]*Department of Chemistry, Northwestern University, Evanston, IL 60208, USA*

TaCo$_2$Te$_2$ is recently reported to be an air-stable, high mobility Van der Waals material with probable magnetic order. Here we investigate the scaling behavior of its magnetoresistance. We measured both the longitudinal ($\rho_{xx}$) and Hall ($\rho_{xy}$) magnetoresistivities of TaCo$_2$Te$_2$ crystals in magnetic fields parallel to the *c*-axis and found that the magnetoresistance violates the Kohler's rule $MR \sim f[H/\rho_0]$ while obeying the extended Kohler's rule $MR \sim f[H/(n_T\rho_0)]$, where $MR \sim [\rho_{xx}(H)-\rho_0]/\rho_0$, $H$ is the magnetic field, $n_T$ is a thermal factor, $\rho_{xx}(H)$ and $\rho_0$ are the resistivities at $H$ and zero field, respectively. While deviating from those of the densities of electrons ($n_e$) and holes ($n_h$) obtained from the two-band model analysis of the magnetoconductivities, the temperature dependence of $n_T$ is close to that of the Hall carrier densities $n_H$ calculated from the slopes of $\rho_{xy}(H)$ curves at low magnetic fields, providing a new way to obtain the thermal factor in the extended Kohler's rule.

*Corresponding author, xiao@anl.gov or zxiao@niu.edu

## I. Introduction

Magnetic field dependent resistance is a long-studied property of materials [1-5]. The magnetic-field-induced resistance change of a material, termed as magnetoresistance (MR) [1], can have large values based on intrinsic sample properties, such as colossal MR (CMR) [2] and giant MR (GMR) [3] in magnetic materials, and extremely large MR (XMR) in non-magnetic materials [4]. Kohler developed a rule in 1938 to describe the scaling behavior of magnetoresistance in metals [5]. This rule states that MR will scale as a function of magnetic field $H$ divided by zero-field resistivity $\rho_0$ of a material, such that $MR = f[H/\rho_0]$, where $MR = [\rho_{xx}(H)-\rho_0]/\rho_0$ and $\rho_{xx}(H)$ is the longitudinal resistivity at $H$. Kohler's rule has since been applied to other non-magnetic materials such as semimetals [6-11] and superconductors (in the normal state) [12]. Violations of Kohler's rule have been widely reported [13,14-16] and are used as evidence for phase transitions [6-8]. An extended Kohler's rule [13] has recently been developed and shown to resolve these violations for a variety of materials by introducing a thermal factor $n_T$, such that $MR = f[H/(n_T\rho_0)]$. In general, $n_T$ is associated with the Fermi level and dispersion relation of the material. Thus, it could be challenging to derive its temperature dependence analytically while it can be obtained experimentally from the scaling behavior. Here, we demonstrate that the extended Kohler's rule is valid in $TaCo_2Te_2$ and the thermal factor $n_T$ has nearly the same temperature dependence of the charge carrier density $n_H$ derived from the Hall resistivity at low magnetic fields, providing a new way to obtain $n_T$.

$TaCo_2Te_2$ is recently reported as an air-stable high mobility Van der Waals material with probable magnetic order, compensated electron-hole densities, and exfoliable down to a few layers [17]. The atomic structure is reported as orthorhombic with a known Peierls distortion in the Ta-Co chains [18]. It has also been reported to have a complex Fermi surface with multiple Fermi pockets, eightfold fermions and fourfold degenerate van Hove singularity [19] as well as large intrinsic nonlinear Hall effect [20]. Resistivity measurements show non-saturating magnetoresistance, with *MR* being as large as $3.72\times10^3$ % at $T = 3$ K and $H = 9$ T [21]. Annealing has shown to have effects on the magnetotransport and magnetic properties of the material, such as carrier mobilities that have a direct effect on the magnitude of magnetoresistance [21]. The magnetoresistances in $TaCo_2Te_2$ have been analyzed using the two-band model, yielding the density and mobility of the charge carriers [17,21]. Kohler's rule plots have also been tested for magnetoresistances in both as-grown and annealed crystals measured at temperatures up to $T = 75$ K and violations could be seen [21]. In this work we probe Kohler's rule for magnetoresistances obtained at temperatures up to $T = 200$ K, enabled by high precision ac resistivity measurements that yield low-noise *MR*s down to $10^{-2}$ %. The extension of the temperature limit to higher values, which increases carrier densities due to enhanced thermal excitation, enables us to evidently demonstrate the violation of the Kohler's rule, test the newly developed extended Kohler's rule and compare the temperature dependence of the thermal factor $n_T$ in the extended Kohler's rule, the charge carrier densities $n_e$ and $n_h$ derived from the two-band model analysis, and $n_H$ from the low field Hall resistivity.

## II. Experimental Details

Single crystals $TaCo_2Te_2$ were grown by chemical vapor transport method with $I_2$ as the transport agent [19]. A mixture of Ta powder (purity 99.9%), Co pieces (99.99%), and Te shot (99.99%) were prepared with a molar ratio of Ta: Co: Te = 1: 1:2 and sealed in a quartz tube under high vacuum, which was then placed in a horizon two-zone tube furnace and maintained under a temperature gradient of 950–850 °C. After 2 weeks, shiny single crystals with a typical size of about 1 mm × 2 mm × 0.1 mm were obtained. More characterizations including x-ray diffraction can be found in Ref.19.

Both single crystal and exfoliated samples were prepared. Contacts were made to an exfoliated sample using standard photolithography techniques. The sample was first mechanically exfoliated using Scotch tape. It was then deposited on a $SiO_2$ substrate. Contacts were made using photolithography micropatterning. After photolithography, 200 nm of gold following a 5 nm adhesion layer of titanium were deposited using magnetron sputtering method. The distance between $R_{xx}$ and $R_{xy}$ contacts are designed to be 10 μm and 20 μm respectively and confirmed with optical microscopy. The thickness of the exfoliated sample was measured using a Bruker Fast Scan AFM to be ~155 nm. Electrical leads for single crystal samples (with a thickness of ~100 μm) were gold wires glued to the crystal using silver epoxy H20E.

We took data on longitudinal and Hall resistances $R_{xx}(H)$ and $R_{xy}(H)$ using the Electrical Transport Option (ETO) of Quantum Design PPMS®. Low-frequencies (33.57 Hz for $R_{xx}$ and 21.36 Hz for $R_{xy}$) ac current of 1 mA was used and flowed in the *ab* plane. The magnetic field was applied along the *c*-axis of the crystal, which is perpendicular to the plane of the current and voltage contacts. We symmetrize the data using positive and negative fields, $R_{xx} = [R_{xx}(+H) + R_{xx}(-H)]/2$ and $R_{xy} = [R_{xy}(+H) - R_{xy}(-H)]/2$ to eliminate contributions from potential contact misalignment and non-uniform current distribution. The resistivities are calculated as $\rho_{xx} = \frac{R_{xx}wd}{l}$ and $\rho_{xy} = R_{xy}d$, where *d*, *w*, and *l* are the thickness, width of the sample, and separation between voltage contacts, respectively.

We measured both single crystal and exfoliated samples. The single crystal sample shows similar trends to the data presented in this paper. We choose to present data from the exfoliated sample due to noise in the low voltage signal of the larger sample. Exfoliation of the material allows us to decrease the thickness and thus increase the resistance of the sample to obtain a larger voltage signal, reducing the effects of noise on the data.

### III. Results and Discussion

Figure 1 presents the temperature dependence of the sample as it undergoes zero-field cooling. It shows a metal-like behavior without apparent transitions. The sample has a residual resistivity ratio (RRR) of ~48, suggesting good sample quality. This value falls between RRR = 17 and 73 reported in Ref.17 and Ref.21 for the as-grown crystals, respectively while being exactly the same as that of the annealed crystal [21]. As presented below, these RRR values are indeed direct indicators of the sample properties such as the amplitudes of the magnetoresistances. That is, MRs in our sample are expected to be close to those of the annealed crystals [21] while being larger than those reported in Ref.17 and smaller than those reported in Ref.21 for as-grown crystals.

In previous reports $\rho_0(T)$ in the low-temperature region ($T \leq 30$ K) was found to follow a $\rho_0(T) \sim T^4$ relationship, which differs from the $\rho_0(T) \sim T^2$ expected for pure electron–electron scattering or a $\rho_0(T) \sim T^5$ expected for the electron–phonon scattering and was attributed to interband electron-phonon scattering [17, 21]. As indicated by the blue dashed line in Fig.1, $\rho_0(T)$ in our sample can be fitted with $\rho_0(T) \sim T^4$ for temperatures up to $T \approx 30$ K. On the other hand, a power-law relationship with an exponent of 2.7 is revealed in the logarithmic plot, as presented in the inset. Thus, power-law fittings may be oversimplified or unreliable in uncovering the scattering mechanisms. We tried to fit our $\rho_0(T)$ with the well-known Bloch-Gruneisen function [22-28]:

$$\rho(T) = \rho_0(0) + A \left(\frac{T}{\theta_D}\right)^n \int_0^{\frac{\theta_D}{T}} \frac{x^n}{(e^x-1)(1-e^{-x})} dx, \qquad (1)$$

where $A$ is a weighting parameter, $T$ is temperature, $\theta_D$ is the Debye temperature, $\rho_0(0)$ is the resistivity plateau at low temperatures and $n$ is a scattering-dependent parameters, with $n = 2, 3, 5$ corresponding to electron-electron, electron-magnon, and electron-phonon scattering, respectively [28]. As presented in Fig.1 as a red solid line, $\rho_0(T)$ in a much wider range (up to $T \leq 100$ K) can be well described by Eq.1 with $n = 5$, along with $A = 430$, $\theta_D = 188$ K and $\rho_0(0) = 2.25$ µΩcm. A fit with $n$ as a free parameter also yields $n = 4.78$. Thus, we conclude that $\rho_0(T)$ is governed by electron-phonon scattering. At high temperatures ($T > 100$ K) the measured resistivities are lower than those expected from Eq.1. As we will reveal below from the violation of Kohler's rule of magnetoresistance, this deviation is caused by thermally induced increase in the charge carrier densities, which is not accounted for in the Bloch-Gruneisen function. That is, such a deviation may indicate the violation of Kohler's rule of magnetoresistance, which is the focus of this work.

We measured $R_{xx}(H)$ and $R_{xy}(H)$ in intervals of $\Delta T = 5$ K at $T \leq 150$ K and then increased to large temperature intervals as the magnetoresistance quickly decreases at higher and higher temperatures. Data was taken up to $T = 300$K but magnetoresistance is negligible at $T > 200$ K. As such, we will focus on data only up to $T = 200$ K, as presented in Figs.2(a) and 2(b) for the $\rho_{xx}(H)$ curves at a few fixed temperatures and $\rho_{xx}(T)$ curves at a few fixed magnetic fields, respectively. We obtained $\rho_{xx}(T)$ curves from the $\rho_{xx}(H)$ curves rather than directly measured them by sweeping temperature to avoid nonequilibrium temperature effects. As expected, $\rho_{xx}(H)$ data in Fig.2(a) show that the magnetic field has stronger effects on the resistivity at lower temperatures, e. g., the $\rho_{xx}(H)$ curve at 3K crosses those at $T = 25$ K and 35 K at high fields. The $\rho_{xx}(T)$ curves in Fig.2(b) are also consistent with those reported in Ref.[21], including the turn-on temperature behavior at $H \geq 5$ T.

We present quantitative analysis of the magnetoresistance in Fig. 3, with the calculated $MR(H)$ and $MR(T)$ in Figs.3(a) and 3(b), respectively. We use logarithmic scale to exhibit the data of small values, particularly the MR curves obtained at high temperatures. Furthermore, $MR(H)$ plots in logarithmic scale enable to show that the curves are parallel to each other, as expected from the Kohler's rule, which suggests that a multiplier of $1/\rho_0$ to $H$ ($x$ axis) could cause them to overlap or collapse onto the same curve [13].

We see large $MR$ gain at low temperatures, reaching up to $MR$ of $\sim 8.33 \times 10^2$ % at $T = 3$ K and $H = 9$ T. This value lies in the middle of those ($2.63 \times 10^2$ %-$3.72 \times 10^3$ %) reported for the as-grown crystals [17,21] while being close to that ($5.71 \times 10^2$ %) of the annealed crystals [21], consistent with their relative RRR values. Figure 3(a) indicates that $MR(H)$ does not follow a simple power law, similar to that observed in multiband semimetals [9,10,13,29]. Figure 3(b) reveals that the $MR(T)$ is very sensitive to temperature at $T > 20$ K, with its value down to $MR \approx 0.5$ % at $T = 200$ K and $H = 9$ T, while reaching a plateau at $T \leq 10$ K.

Kohler's rule tells us that the curves when plotted as $MR \sim H/\rho_0$ should collapse onto each other, showing an invariance in MR scaling with temperature. We can see from Fig.3(c) that Kohler's rule is followed very well for $T \leq 50$ K whereas its violation can be clearly identified for the curve at $T = 75$ K and becomes more prominent with increasing temperature. On the other hand, those curves at high temperatures are still parallel to each other, suggesting that an additional temperature dependent scaling factor $n_T$ are needed to collapse all $MR(H/\rho_0)$ to a single curve, leading to the extended Kohler's rule $MR \sim f[H/(n_T\rho_0)]$ proposed in Ref. 13 and verified by others [29-32]. Following the same procedures used in Ref.13 we use $n_T = 1$ for MR at $T = 200$ K, i.e., this scaling factor is normalized to scale all curves to the $T = 200$ K curve. As shown in Figs.3(e) and 3(f), all the curves $MR[H/(n_T\rho_0)]$ indeed overlap each other, confirming the applicability of the extended Kohler's rule in TaCo$_2$Te$_2$. The corresponding $n_T$ values obtained at various temperatures are presented in Fig.4(a), which show significant temperature dependence at $T > 50$ K while being nearly constant at $T \leq 50$ K. As indicated by the red line, it can be fitted with a power-law relationship $n_T \sim T^\nu$ with $\nu = 2.2$, which is very close to that ($\nu = 2$) reported for TaP, where $n_T$ can be theoretically described by the temperature dependence of the charge carrier densities [13]. That is, $n_T$ of TaCo$_2$Te$_2$ is probably dominated by thermally-induced change in its charge carrier densities. However, TaCo$_2$Te$_2$ is a multiband semimetal with more complicated bandstructure [17,19] than that of TaP, hindering the calculation of its temperature dependent charge carrier densities. Below we will compare it with the temperature dependence of carrier densities derived from the same sets of magnetoresistivity data using various analysis approaches and demonstrate that $n_T$ has in fact similar temperature dependence as the Hall carrier density $n_H$ determined from the low-field Hall resistivities.

In the literature two-band model fittings of longitudinal and/or Hall magnetoresistivities/conductivities have been widely used to obtain the electron and hole carrier densities $n_e$ and $n_h$ [33-36]. In fact, they have also been applied on TaCo$_2$Te$_2$ [17,21], yielding $n_e$ and $n_h$ whose values decrease significantly with increasing temperature at $T \leq 75$ K [21], which differs from temperature-insensitive $n_T$ in that temperature regime as presented in Fig.4. On one hand, this inconsistence may arise from the two-band model fitting approach itself, which could result in capricious outcomes in TaCo$_2$Te$_2$ that is not a two-band system, as demonstrated in our recent work on ZrSiSe [36]. On the other hand, it might occur simply because the data are from different samples, since carrier densities can depend strongly on the synthesis conditions of the crystals [21]. For a reliable comparison on the temperature dependence of $n_e$ and $n_h$ to that of $n_T$, we also conduct two-band model analysis on magnetoconductivities $\sigma_{xx}$ and $\sigma_{xy}$ of our TaCo$_2$Te$_2$ sample. The magnetoconductivities are converted from the measured magnetoresistivities $\rho_{xx}$ and $\rho_{xy}$ via $\sigma_{xx} = \frac{\rho_{xx}}{\rho_{xx}^2 + \rho_{xy}^2}$ and $\sigma_{xy} = \frac{\rho_{xy}}{\rho_{xx}^2 + \rho_{xy}^2}$. By limiting the discussions in Ref.13 on multi-band systems to $i$ = 1 and 2 and assigning $n_1 = n_h$, $\mu_1 = \mu_h$ for holes and $n_2 = -n_e$, $\mu_2 = -\mu_e$ for electrons, we obtain the following expressions:

$$\sigma_{xy} = eH \left\{ \frac{n_h \mu_h^2}{1+(\mu_h H)^2} - \frac{n_e \mu_e^2}{1+(\mu_e H)^2} \right\} \tag{2}$$

$$\sigma_{xx} = e \left\{ \frac{n_h \mu_h}{1+(\mu_h H)^2} + \frac{n_e \mu_e}{1+(\mu_e H)^2} \right\}, \tag{3}$$

where $e$ is the charge of the electron, $H$ is the magnetic field, $n_h$, $\mu_h$, $n_e$, $\mu_e$ are carrier densities and mobilities of holes and electrons, respectively. We then simultaneously fit the experimental $\sigma_{xx}(H)$ and $\sigma_{xy}(H)$ curves with Eqs.(2) and (3), as shown in Fig.5 for $T$ = 3 K and 100 K as examples. While the two-band model does describe the data well at high temperatures, the fitting curves deviate significantly from the measured ones at low temperatures, consistent with that reported in Ref.[17]. At $T$ = 3 K, we obtain $n_h$ = 3.07 × $10^{26}$ m$^{-3}$, $n_e$ = 3.45 × $10^{26}$ m$^{-3}$, $\mu_h$ = 0.462 m$^2$V$^{-1}$s$^{-1}$, and $\mu_e$ = 0.364 m$^2$V$^{-1}$s$^{-1}$. These values are on the same order of magnitude as those presented in previous works [17,21]. We see that the density of electrons is larger than that of holes in our sample, in agreement with initial report [17]. As shown in Fig.4(b) for $n_e$ and $n_h$ obtained at temperatures up to $T$ = 200 K, their values are the same at $T$ > 50 K while have small difference at $T \leq$ 50 K, consistent with electron-hole compensation claimed previously [17]. Furthermore, their temperature dependence indeed follows a similar trend as that in Ref.21 at $T$ < 75 K, though the temperature sensitivity in our sample is weaker. At $T$ > 75 K, the values of $n_e$ and $n_h$ do become larger with increasing temperature, similar to that of $n_T$. However, it increases only by a factor of ~1.7, which is significantly smaller than the factor of ~2.5 for $n_T$ when the temperature is increased from $T$ = 75 K to $T$ = 200 K. Both this weaker temperature sensitivity and the nonmonotonic temperature dependence indicate that the temperature dependence of the $n_e$ and $n_h$ obtained from two-band model analysis of the magnetconductivities do not reflect that of the thermal factor $n_\text{T}$.

On the other hand, we can derive the thermal factor for a multi-band material at magnetic fields satisfying $\mu_i H \ll 1$, which can be expressed as $n_\text{T} = e \frac{[\sum_i n_i \mu_i]^{3/2}}{[\sum_i n_i \mu_i^3]^{1/2}}$, where $n_i$ and $\mu_i$ are the density and mobility of the charge carriers in the $i^\text{th}$ band [13]. Namely, the thermal factor $n_T$ contains contributions from the temperature dependences of the carrier densities and mobilities of all bands. In the simplest case that the densities or mobilities from different bands have the same or similar temperature dependence, i.e., $n_i = n_i^0 f_n(T)$ and $\mu_i = \mu_i^0 f_\mu(T)$, we can have $n_\text{T} = e \frac{[\sum_i n_i^0 \mu_i^0]^{3/2}}{[\sum_i n_i^0 (\mu_i^0)^3]^{1/2}} f_n(T) \sim f_n(T)$. That is, $n_\text{T}$ is governed by the temperature dependence of the carrier density. In this case the Hall resistivity can also be reduced to $\rho_{xy} = \frac{\Sigma_i(en_i\mu_i^2 H)}{[\Sigma_i(en_i\mu_i)]^2}$, i.e., $\rho_{xy}(H)$ curves are linear at low fields. Together with the definition of Hall carrier density $n_\text{H}$, i.e., $\rho_{xy} = \frac{H}{en_\text{H}}$, we obtain $n_\text{H} = [\Sigma_i(n_i^0 \mu_i^0)]^2 / \Sigma_i[n_i^0(\mu_i^0)^2] f_n(T) \sim f_n(T)$. That said, the Hall carrier density $n_\text{H}$ determined from the slope of the linear $\rho_{xy}(H)$ curve at low fields is expected to have similar temperature dependence as that of $n_\text{T}$.

Figure 6(a) presents $\rho_{xy}(H)$ curves calculated from the $R_{xy}(H)$ curves obtained simultaneously with $R_{xx}(H)$ curves used to derive the $\rho_{xx}(H)$ curves in Fig.2(a). While they are linear at high temperatures ($T \geq 100$ K), the $\rho_{xy}(H)$ curves become obviously non-linear at high magnetic fields at low temperatures ($T \leq 75$K), resembling those reported in other multi-band semimetals [10,11,25,30,36]. On the other hand, Figure 6(b) shows that the $\rho_{xy}(H)$ curves are linear at temperatures down to $T = 3$ K at $H \leq 0.2$ T. From the estimated mobilities from the two-band model analysis, we see that $\mu_i H \ll 1$ is satisfied in this field range, consistent with the above theoretical consideration. From the slopes of the low-field $\rho_{xy}(H)$ curves we obtain the Hall carrier densities $n_\text{H}$, which are presented in Fig.4(b). Similar to $n_\text{T}$, $n_\text{H}$ becomes more temperature-sensitive with increasing temperatures. The small dip-like feature around $T$ = 20 K in the $n_\text{H}(T)$ curve is likely due to fitting errors, since Kohler's rule is followed closely in this regime, i.e., carrier densities are expected to be temperature-insensitive. For a better comparison with the temperature dependence of $n_\text{T}$, we define a normalized Hall carrier density $n_\text{H}^* = n_\text{H}(T)/n_\text{H}(200\text{K})$, i.e., dividing the temperature-dependent Hall carrier densities by its value at $T$ = 200 K, and plot it together with $n_\text{T}$ in Fig.4(a). We see that $n_\text{H}^*$ follows nearly the same temperature dependence of $n_\text{T}$. As indicated by the violet line in Fig.4(a), the temperature dependence of $n_\text{H}^*$ also follows a power-law relationship $n_\text{H}^* \sim T^\nu$ with $\nu$ = 2.1, closely resembling that ($\nu$ = 2.2) of $n_\text{T}$. Figures 3(g) and 3(h) further demonstrate that scalings of similar quality can be achieved when $n_\text{T}$ in the extended Kohler's rule is replaced with $n_\text{H}^*$. These results clearly show that the temperature sensitive carrier densities are the dominant contributors to the violation of Kohler's rule in TaCo$_2$Te$_2$. They also demonstrate an alternative way to obtain the thermal factor in the extended Kohler's rule. On the other hand, Figure 4(b) shows that the values of $n_\text{H}$ are 10~20 times larger than those of $n_\text{e}$ and $n_\text{h}$. While the two-band model analysis may only yield an estimate of the values of $n_\text{e}$ and $n_\text{h}$, such a large difference is most likely caused by the definition of $n_\text{H}$. As discussed above, it can be expressed as $n_\text{H} = [\Sigma_i(n_i\mu_i)]^2/\Sigma_i(n_i\mu_i^2)$, indicating that its values can differ significantly from those of the true carrier densities in the material, though it is related to and can even be proportional to them, as evidently demonstrated in Fig.S2 and its caption in Ref.13 for a compensated two-band material with various ratio of $\mu_h/\mu_e$. That is, the absolute values of $n_\text{H}$ may be unable to provide reliable information on the true carrier densities but its temperature dependence can reflect that of the thermal factor in the extended Kohler's rule.

**IV. Conclusions**

In summary, we investigated the magnetoresistance of TaCo$_2$Te$_2$ at temperatures up to 200 K and showed it has a clear violation to Kohler's rule at high temperatures. We demonstrated the validity of the extended Kohler's rule in TaCo$_2$Te$_2$ and compared the temperature dependence of the derived thermal factor $n_\mathrm{T}$ with those of the electron and hole densities $n_\mathrm{e}$ and $n_\mathrm{h}$ estimated from two-model analysis as well as the Hall carrier density $n_\mathrm{H}$ obtained from the low-field Hall resistivities. We show that $n_\mathrm{T}$ and $n_\mathrm{H}$ have similar temperature dependence, highlighting the role of temperature-sensitive carrier densities in the violation of Kohler's rule of magnetoresistance and providing a new experimental way to obtain the thermal factor in the extended Kohler's rule.

**Acknowledgement**

Resistivity measurements were supported by the U.S. Department of Energy, Office of Science, Basic Energy Sciences, Materials Sciences and Engineering. S.E.P & Z.L.X acknowledge support from the National Science Foundation grant# DMR-1901843. L.L. & B.S. were supported by the National Natural Science Foundation of China (NSFC) grant # U213010013, Natural Science Foundation of Guangdong Province grant# 2022A1515010035, Guangzhou, Basic and Applied Basic Research Foundation grant# 202201011798, and the open research fund of Songshan Lake materials Laboratory grant # 2021SLABFN11. Work performed at the Center for Nanoscale Materials, a U.S. Department of Energy Office of Science User Facility, was supported by the U.S. DOE, Office of Basic Energy Sciences, under Contract No. DE-AC02-06CH11357.

**References**

[1] J. M. Ziman, Electrons and phonons: The theory of transport phenomena in solids (Cambridge University Press, Cambridge, United Kingdom, 2001).

[2] A. P. Ramirez, R. J. Cava, and J. Krajewski, Colossal magnetoresistance in Cr-based chalcogenide spinels, Nature (London) **386**, 156 (1997).

[3] J. M. Daughton, GMR Applications, J. Magn. Magn. Mater. **192**, 334 (1999).

[4] R. Niu, W. Zhu, Materials and possible mechanisms of extremely large magnetoresistance: A review, J. Phys.: Condens. Matter. **34**, 13001 (2021).

[5] M. Kohler, Zur magnetischen widerstandsänderung reiner metalle, Ann. Phys. (Berlin) **424**, 211 (1938).

[6] Y. L. Wang, L. R. Thoutam, Z. L. Xiao, J. Hu, S. Das, Z. Q. Mao, J. Wei, R. Divan, A. Luican-Mayer, G. W. Crabtree, and W. K. Kwok, Origin of the turn-on temperature behavior in $WTe_2$, Phys. Rev. B **92**, 180402(R) (2015).

[7] Q. L. Pei, W. J. Meng, X. Luo, H. Y. Lv, F. C. Chen, W. J. Lu, Y. Y. Han, P. Tong, W. H. Song, Y. B. Hou, Q. Y. Lu, and Y. P. Sun, Origin of the turn-on phenomenon in $Td$-$MoTe_2$, Phys. Rev. B **96**, 075132 (2017).

[8] N. H. Jo, Y. Wu, L.-L. Wang, P. P. Orth, S. S. Downing, S. Manni, D. X. Mou, D. D. Johnson, A. Kaminski, S.L. Bud'ko, and P. C. Canfield, Extremely large magnetoresistance and Kohler's rule in $PdSn_4$: A complete study of thermodynamic, transport, and band-structure properties, Phys. Rev. B **96**, 165145 (2017).

[9] F. Han, J. Xu, A. S. Botana, Z. L. Xiao, Y. L. Wang, W. G. Yang, D. Y. Chung, M. G. Kanatzidis, M. R. Norman, G. W. Crabtree, and W. K. Kwok, Separation of electron and hole dynamics in the semimetal LaSb, Phys. Rev. B **96**, 125112 (2017).

[10] J. Song, J. Wang, Y. H. Wang, L. Zhang, M. Song, Z. H. Li, L. Cao, D. Y. Liu, Y. M. Xiong, Kohler's rule and anisotropic Berry-phase effect in nodal-line semimetal ZrSiSe, J. Appl. Phys. **131**, 065106 (2022).

[11] J. H. Du, Z. F. Lou, S. N. Zhang, Y. X. Zhou, B. J. Xu, Q. Chen, Y. Q. Tang, S. J. Chen, H. C. Chen, Q. Q. Zhu, H. D. Wang, J. H. Yang, Q. S. Wu, O. V. Yazyev, and M. H. Fang, Extremely large magnetoresistance in the topologically trivial semimetal α-WP$_2$, Phys. Rev. B **97**, 245101 (2018).

[12] L. Forro, K. Biljakovic, J. R. Cooper and K. Bechgaard, Magnetoresistance of the organic superconductor bistetramethyltetraselenafulvalenium perchlorate [(TMTSF)$_2$ClO$_4$]: Kohler's rule, Phys. Rev. B **29**, 2839 (1984).

[13] J. Xu, F. Han, T.-T. Wang, L. R. Thoutam, S. E. Pate, M. D. Li, X. F. Zhang, Y.-L. Wang, R. Fotovat, U. Welp, X. Q. Zhou, W.-K. Kwok, D. Y. Chung, M. G. Kanatzidis, and Z.-L. Xiao, Extended Kohler's rule of magnetoresistance, Phys. Rev. X **11**, 041029 (2021).

[14] A. Wang, D. Graf, A. Stein, Y. Liu, W. Yin, and C. Petrovic, Magnetotransport properties of MoP$_2$, Phys. Rev. B **96**, 195107 (2017).

[15] I. A. Leahy, Y.-P. Lin, P. E. Siegfried, A. C. Treglia, J. C. W. Song, R. M. Nandkishore, and M. Lee, Nonsaturating large magnetoresistance in semimetals, Proc. Natl. Acad. Sci. U.S.A. **115**, 10570 (2018).

[16] P. Cheng, H. Yang, Y. Jia, L. Fang, X. Y. Zhu, G. Mu, and H.-H. Wen, Hall effect and magnetoresistance in single crystals of NdFeAsO$_{1-x}$F$_x$, Phys. Rev. B **78**, 134508 (2008).

[17] R. Singha, F. Yuan, G. M. Cheng, T. H. Salters, Y. M. Oey, G. V. Villalpando, M. Jovanovic, N. Yao, and L. M. Schoop, TaCo$_2$Te$_2$: An air-stable, high mobility Van der Waals material with probable magnetic order, Adv. Funct. Mater. **32**, 2108920 (2022).

[18] W. Tremel, TaNi$_2$Te$_2$, A Novel Layered Telluride, and TaCo$_2$Te$_2$, a Structural Variant with Peierls Distortion, Angew. Chem., Int. Ed. **31**, 217 (1992).

[19] H. T. Rong, Z. Q. Huang, X. Zhang, S. Kumar, F. Y. Zhang, C. C. Zhang, Y. Wang, Z. Y. Hao, Y. Q. Cai, L. Wang, C. Liu, X. M. Ma, S. Guo, B. Shen, Y. Liu, S. T. Cui, K. Shimada, Q. S. Wu, J. H. Lin, Y. G. Yao, Z. W. Wang, H. Xu, and C. Y. Chen, Realization of practical eightfold fermions and fourfold van Hove singularity in TaCo$_2$Te$_2$, npj Quantum Mater. **8**, 29 (2023).

[20] F. Mazzola, B. Ghosh, J. Fujii, G. Acharya, D. Mondal, G. Rossi, A. Bansil, D. Farias, J. Hu, A. Agarwal, A. Politano, and I. Vobornik, Discovery of a magnetic Dirac system with a large intrinsic nonlinear Hall effect, Nano Lett., **23**, 902 (2023).

[21] L. S. Wang, J. J. Tian, C. Y. Kang, H. Y. Gu, R. Pang, M. N. Shen, L. M. She, Y. H. Song, X. S. Liu, and W. F. Zhang, Effect of post-annealing on magnetotransport and magnetic properties of TaCo$_2$Te$_2$ single crystals, Inorg. Chem. **61**, 18899 (2022).

[22] F. Bloch, Zum elektrischen widerstandsgesetz bei tiefen temperaturen, Eur. Phys. J. A **59**, 208 (1930).

[23] E. Grüneisen. Die abhängigkeit des elektrischen widerstandes reiner Metalle von der Temperatur, Ann. Phys. **408**, 530 (1933).

[24] D. Santos-Cottin, A. Gauzzi, M. Verseils, B. Baptiste, G. Feve, V. Freulon, B. Plaçais, M. Casula, and Y. Klein. Anomalous metallic state in quasi-two-dimensional BaNiS$_2$, Phys. Rev. B **93**, 125120 (2016).

[25] O. Pavlosiuk, M. Kleinert, P. Swatek, D. Kaczorowski, and P. Wiśniewski, Surface topology and magnetotransport in semimetallic LuSb, Sci. Rep. **7**, 12822 (2017).

[26] O. Pavlosiuk, P. Swatek, D. Kaczorowski, and P. Wiśniewski, Magnetoresistance in LuBi and YBi semimetals due to nearly perfect carrier compensation. Phys. Rev. B **97**, 235132 (2018).

[27] S. X. Xu, S. Xu, J. P. Sun, B. S. Wang, Y. Uwatoko, T. L. Xia, and J. G. Cheng. Pressure effect on the magnetoresistivity of topological semimetal RhSn, J. Phys.: Condens. Matter. **32**, 355601 (2020).

[28] E. Talantsev, Quantifying the charge carrier Interaction in metallic twisted bilayer graphene superlattices, Nanomaterials **11**, 1306 (2021).

[29] J. Xing, J. Blawat, S. Speer, A. Saleheen, J. Singleton, and R. Y. Jin. Manipulation of the magnetoresistance by strain in topological $TaSe_3$, Adv. Quant. Technol. **5**, 2200094 (2022).

[30] Q. X. Li, B. Wang, N. N. Tang, C. S. Li, E. K. Yi, B. Shen, D. H. Guo, D. Y. Zhong, and H. C. Wang. Signatures of temperature-driven Lifshitz transition in semimetal hafnium ditelluride, Chin. Phys. Lett. **40**, 067101 (2023).

[31] N. K. Karn, M. M. Sharma, and V. P. S. Awana, Weak antilocalization and topological edge states in $PdSn_4$, J. Appl. Phys. **133**, 175102 (2023).

[32] J. M. DeStefano, E. Rosenberg, O. Peek, Y. B. Lee, Z. Y. Liu, Q. N. Jiang, L. Q. Ke, and J.-H. Chu. Pseudogap behavior in charge density wave kagome material $ScV_6Sn_6$ revealed magnetotransport measurements, npj Quant. Mater. **8**, 65 (2023).

[33] J. Hu, S.-Y. Xu, N. Ni, and Z. Q. Mao, Transport of topological semimetals, Annu. Rev. Mater. Res. **49**, 207 (2019).

[34] L. Y. Xing, R. Chapai, R. Nepal, and R. Y. Jin, Topological behavior and Zeeman splitting in trigonal $PtBi_{2-x}$ single crystals, npj Quant. Mater. **5**, 10 (2020).

[35] A. Saleheen, R. Chapai, L. Y. Xing, R. Nepal, D. L. Gong, X. Gui, W. W. Xie, D. P. Young, E. W. Plummer, and R. Y. Jin. Evidence for topological semimetallicity in a chain compound $TaSe_3$, npj Quant. Mater. **5**, 53 (2020).

[36] J. Xu, Y. Wang, S. E. Pate, Y. L. Zhu, Z. Q. Mao, X. F. Zhang, X. Q. Zhou, U. Welp, W.-K. Kwok, D. Y. Chung, M. G. Kanatzidis, and Z.-L. Xiao, Unreliability of two-band model analysis of magnetoresistivities in unveiling temperature-driven Lifshitz transition, Phys. Rev. B **107**, 035104 (2023).

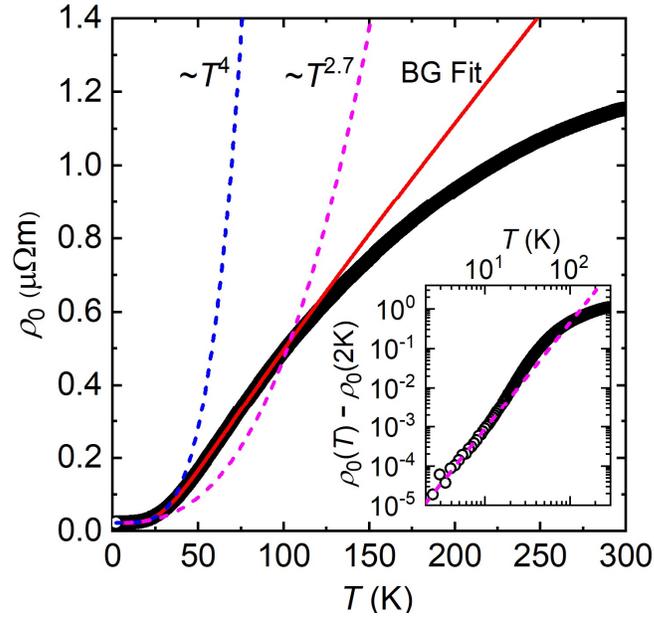

**Fig.1.** Zero-field cooling curve of the exfoliated sample. Symbols are experimental data. Dashed blue line, dashed magenta line and the red solid line present the fits of $\rho_0 \sim T^4$, $\rho_0 \sim T^{2.7}$ and Bloch-Gruneisen function Eq.1, respectively. The inset shows a logarithmic plot of the temperature dependence of the resistivity after subtracting the residual value $\rho_0(2K) = 2.244$ μΩcm at $T = 2$ K. The dashed magenta line represents $\rho_0 \sim T^{2.7}$.

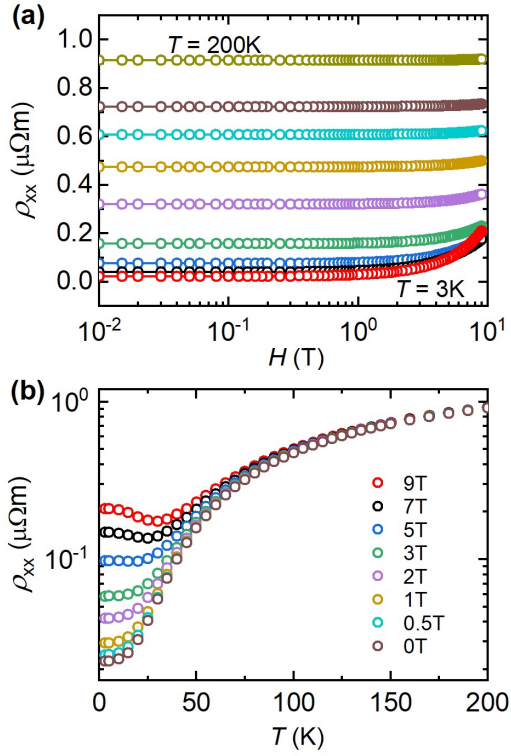

**Fig. 2.** (a) Magnetic field dependence of the longitudinal resistivity, $\rho_{xx}(H)$, at $T$ = 3 K, 25 K, 35 K, 50 K, 75 K, 100 K, 125 K, 150 K, and 200 K (from bottom to top). (b) Temperature dependence of the longitudinal resistivity, $\rho_{xx}(T)$, at various magnetic fields as denoted by the legend.

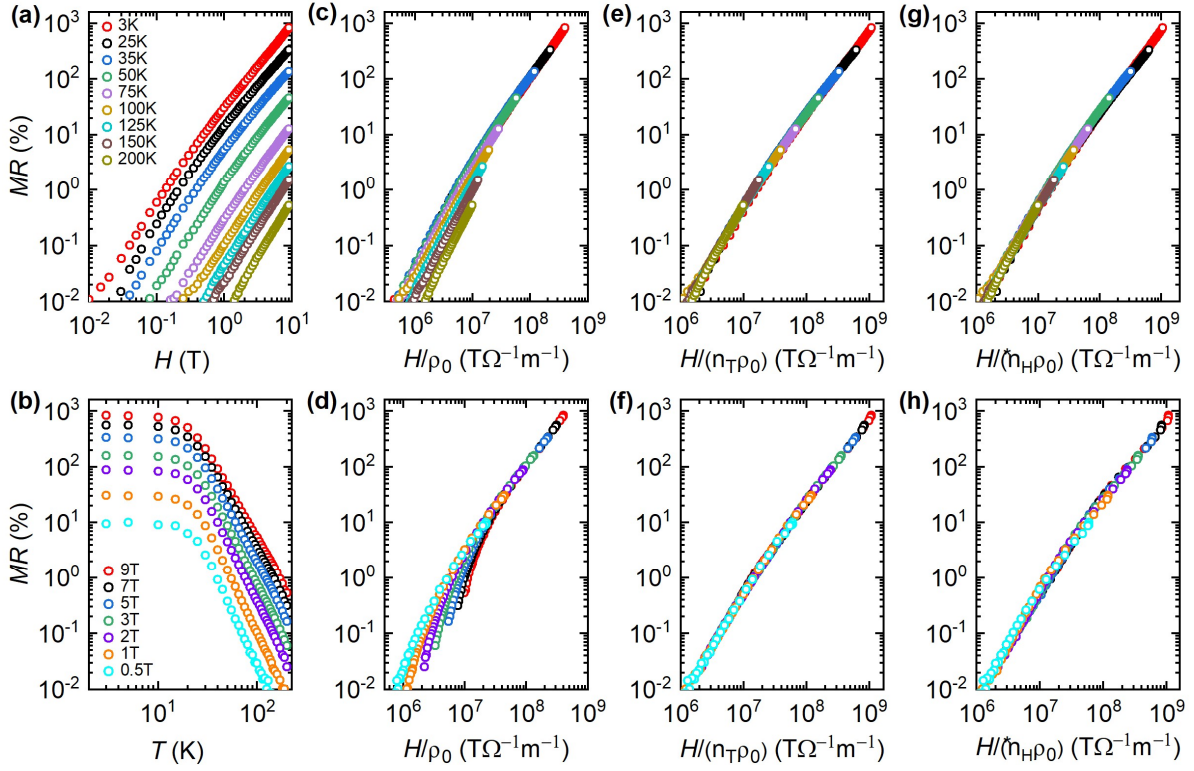

**Fig. 3.** (a) and (b) Magnetic field and temperature dependences of the $MR$ derived from data in Figs. 2(a) and 2(b), respectively. (c) and (d) Kohler's rule plots of the data in (a) and (b), respectively. (e) and (f) Extended Kohler's rule plots of the $MR$ curves in (a) and (b), respectively. The used $n_T$ values are presented in Fig.4(a), where $n_T = 1$ at $T = 200$ K. (g) and (h) Extended Kohler's rule plots of the MR curves in (a) and (b) by replacing $n_T$ with $n_H^*$, where $n_H^* = n_H(T)/n_H(200K)$ is the normalized Hall carrier density, i.e., $n_H^* = 1$ at $T = 200$ K, for direct comparisons with the results in (e) and (f), where $n_H(T)$ is the Hall carrier density obtained from the Hall resistivity at low magnetic fields, as elaborated in Fig.6 and its caption. $n_H^*$ and $n_H$ data are presented in Fig.4(a) and 4(b) respectively.

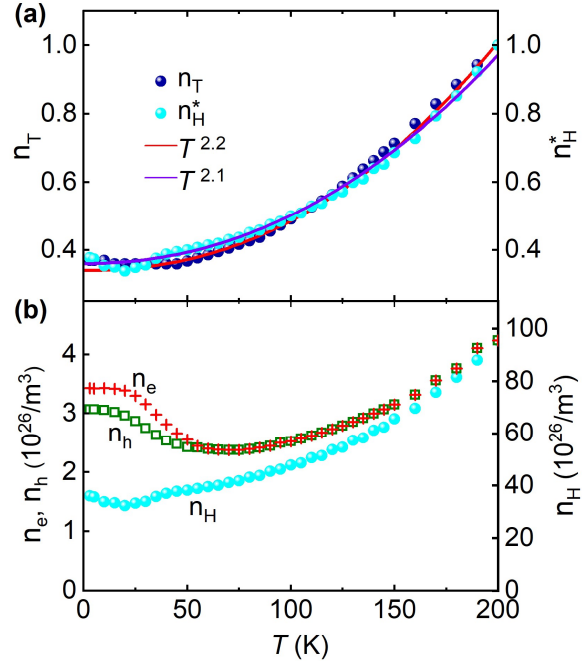

**Fig. 4.** (a) Temperature dependence of the thermal factor $n_T$ and the normalized Hall carrier density $n_H^* = n_H(T)/n_H(200K)$. The red and violet lines represent power-law fittings of $n_T \sim T^{2.2}$ and $n_H^* \sim T^{2.1}$, respectively. (b) Temperature dependence of the Hall carrier density $n_H$ and the electron and hole densities $n_e$ and $n_h$ derived from two-band model analysis.

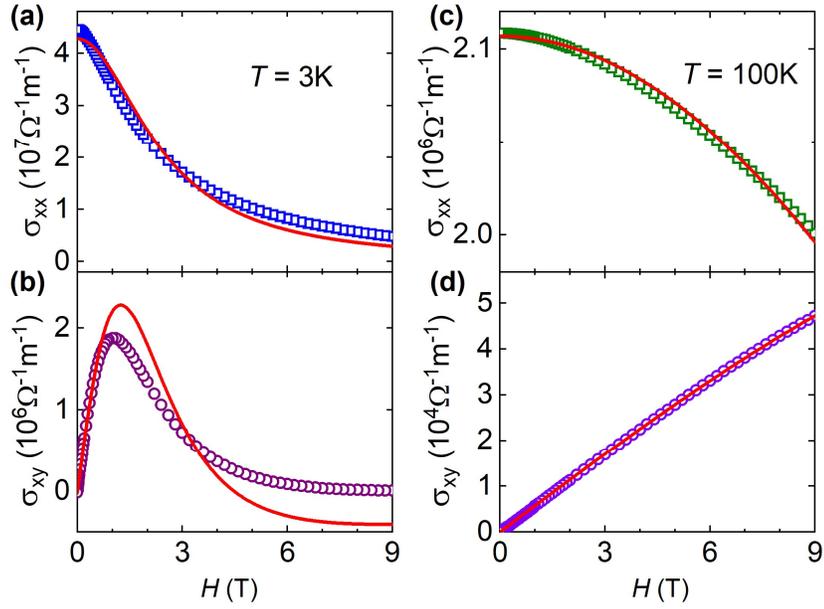

**Fig. 5.** (a) and (b) two-band model fittings of the longitudinal and Hall magnetoconductivities $\sigma_{xx}$ and $\sigma_{xy}$ for $T$ = 3 K, respectively. (c) and (d) the same analysis as those in (a) and (b) but for $T$ = 100 K. Symbols are experimental data and lines are fittings with Eq.(2) and Eq.(3). The derived densities are presented in Fig.4(b).

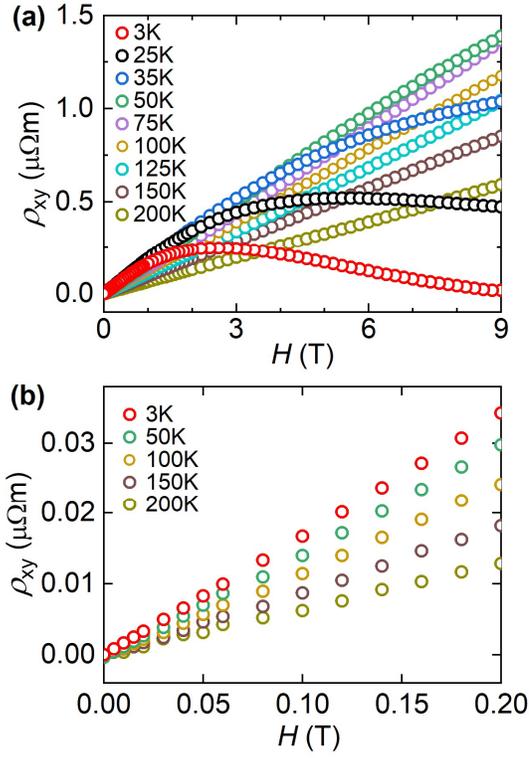

**Fig. 6.** (a) Magnetic field dependence of the Hall resistivity, $\rho_{xy}(H)$, at various temperatures. (b) Expanded view of $\rho_{xy}(H)$ in (a) at magnetic fields up to $H = 0.2$ T. For clarity, only data at $T = 3$ K, 50 K, 100 K, 150 K and 200 K are presented. The linearity of $\rho_{xy}(H)$ allow us to derive the Hall carrier density $n_H = H/(e\rho_{xy})$.